\documentclass[sn-basic, iicol]{sn-jnl}

\jyear{2023}%

\theoremstyle{thmstyleone}%
\theoremstyle{thmstyletwo}%

\theoremstyle{thmstylethree}%

\usepackage{subcaption}
\usepackage{graphicx}
\usepackage{cleveref}
\usepackage[labelsep=space]{caption}
\usepackage{amsmath}
\usepackage{commath}

\Crefname{equation}{Eq.}{Eqs.}
\Crefname{figure}{Fig.}{Figs.}
\Crefname{tabular}{Tab.}{Tabs.}
\Crefname{table}{Tab.}{Tabs.}
\Crefname{section}{Sect.}{Sects.}
\crefname{algocf}{Alg.}{Algs.}

\begin{document}

    \title[Paper's title]{A deep learning-based approach for identifying unresolved questions on Stack Exchange Q\&A communities through graph-based communication modelling}

    \author{\fnm{Hassan} \sur{Abedi Firouzjaei}}\email{hassan.abedi@ntnu.no}

    \affil{\orgdiv{Department of Computer Science}, \orgname{Norwegian University of Science and Technology (NTNU)}, \orgaddress{\city{Trondheim}, \country{Norway}}}

    \abstract{
        In recent years, online question-answering (Q\&A) platforms, such as Stack Exchange (SE), have become increasingly popular for information and knowledge sharing. Despite the vast amount of information available on these platforms, many questions remain unresolved. In this work, we aim to address this issue by proposing a novel approach to identify unresolved questions in SE Q\&A communities. Our approach utilises the graph structure of communication formed around a question by users to model the communication network surrounding it. We employ a property graph model and graph neural networks (GNNs), which can effectively capture both the structure of communication and the content of messages exchanged among users. By leveraging the power of graph representation and GNNs, our approach can effectively identify unresolved questions in SE communities. Experimental results on the complete historical data from three distinct Q\&A communities demonstrate the superiority of our proposed approach over baseline methods that only consider the content of questions. Finally, our work represents a first but important step towards better understanding the factors that can affect questions becoming and remaining unresolved in SE communities.
    }


    \maketitle

    \section{Introduction}

\begin{figure}[t]
        \centerline{\includegraphics[scale=0.33]{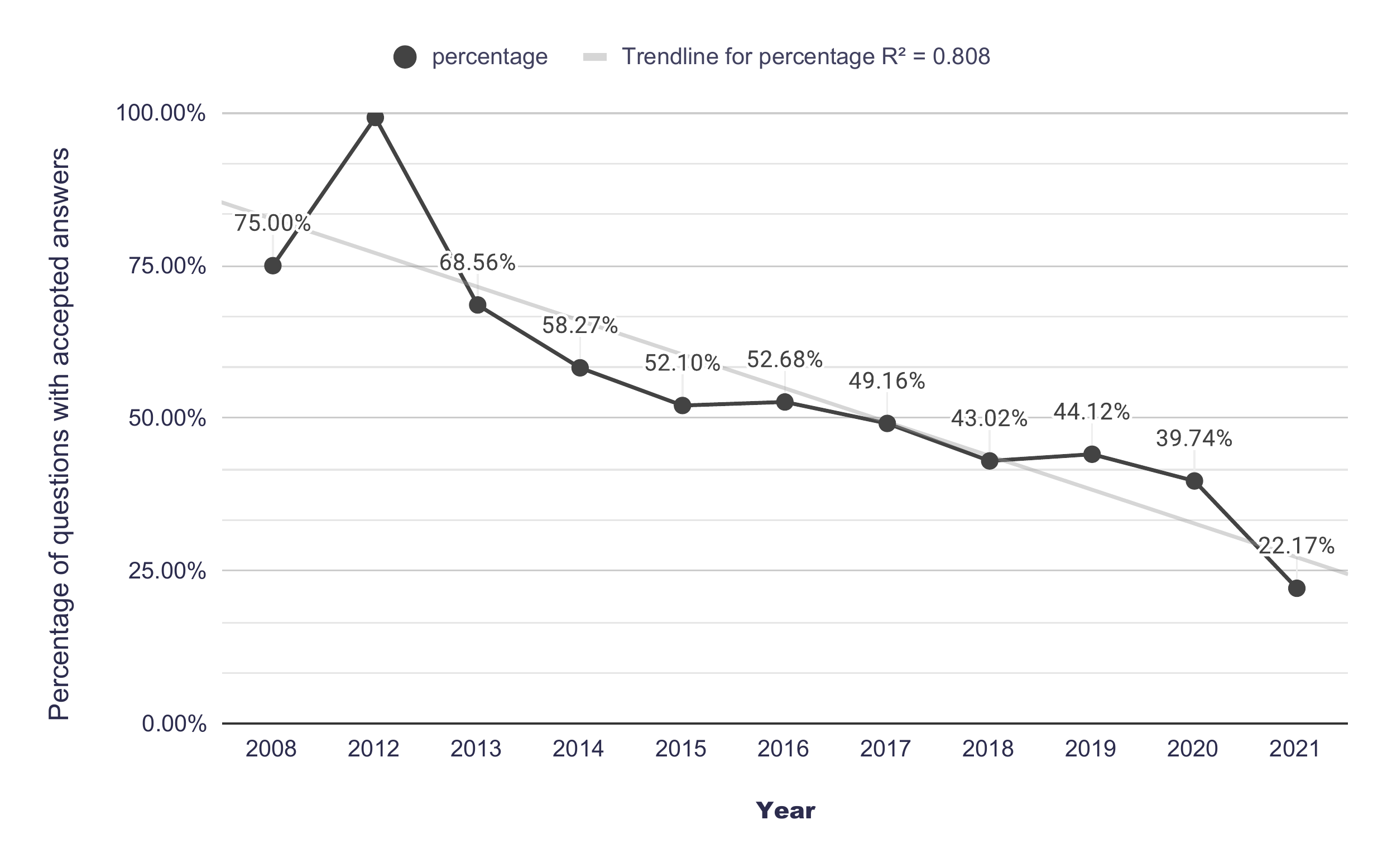}}
        \caption{An illustrative example showing that for the Computer Science SE, like many other communities on the SE platform, the percentage of resolved questions (i.e., questions with accepted answers) followed a decreasing trend over the years}
        \label{fig:motfig1}
    \end{figure}

    Stack Exchange (SE)\footnote{\href{https://stackexchange.com}{https://stackexchange.com}} is a large online platform that hosts over 160 question-answering (Q\&A) communities covering a plethora of topics. At a basic level, on SE, a user asks a question which in turn is answered by others aiming to give the question a satisfactory answer. A question is considered open or unresolved until one of its answers is selected as accepted. This simple process helps facilitate the flow of knowledge from experts to people searching for high-quality information~\citep{AbediFirouzjaei2022}. Nevertheless, many questions on SE may not receive an accepted answer and thus become unresolved. Moreover, a question can be unresolved due to many factors, including the novelty of the topic of the question or its specificity, or even its duplicity. For example, a question can be closed by the community moderators shortly after it was posted because its topic is deemed too narrow.
    A recent study shows that about half of the questions on Stack Overflow, the largest Q\&A community on SE, have yet to get an accepted answer~\citep{Yazdaninia2021}. And the percentage of unresolved questions has been increasing for many communities hosted on the SE platform. ~\Cref{fig:motfig1} shows the same decreasing trend for another community, namely Computer Science SE.

    A graph neural network (GNN) is a deep learning model operating on graph-structured data~\citep{Wu2022}. GNNs have been successfully applied to many tasks, such as node classification, link prediction, and graph classification. They are designed to handle the unique characteristics of graph data, such as variability in size and non-Euclidean structures, primarily by leveraging the graph's topology to propagate information throughout the network. GNNs typically use a message-passing mechanism to aggregate information from neighbouring nodes. This allows the model to learn representations of the graph's structure and node attributes. Moreover, recent advancements in GNNs have resulted in the development of various architectures, such as graph convolutional networks (GCNs)~\citep{Kipf2016} utilising convolutions for graphs~\citep{Wu2022}, graph attention networks (GATs)~\citep{Velickovic2017} based on the attention mechanism~\citep{Chorowski2015, Vaswani2017}, and graph transformer networks (GTNs)~\citep{Yun2019} utilising transformers~\citep{Vaswani2017}. These architectures have achieved state-of-the-art performance on various graph-based tasks such as graph and node classification. They have been applied to problems in chemistry, social network analysis, and computer vision~\citep{Wu2022}. Furthermore, as graph-structured data grows, GNNs are becoming increasingly crucial for various applications.

    The property graph model (PGM) is a flexible and powerful data model used to represent and store graph-structured data~\citep{Bonifati2018}. It is based on a simple yet expressive set of concepts: nodes, edges, and properties. Nodes represent entities in the graph, and edges represent relationships between nodes. Properties are key-value pairs associated with nodes and edges, allowing the graph to store rich, semi-structured data. Moreover, one of the key advantages of the PGM is its ability to handle complex, multi-relational data. For example, it can be used to model social networks, where nodes represent people and edges represent relationships such as friends or family. The properties of nodes and edges can include things like name, age, location, and interests. The PGM also allows easy expression of complex queries and traversals, which is helpful for data analysis. This, in turn, makes the PGM popular for many applications, including graph databases and graph-based machine learning.

\begin{figure*}[th]
        \centerline{\includegraphics[scale=0.4]{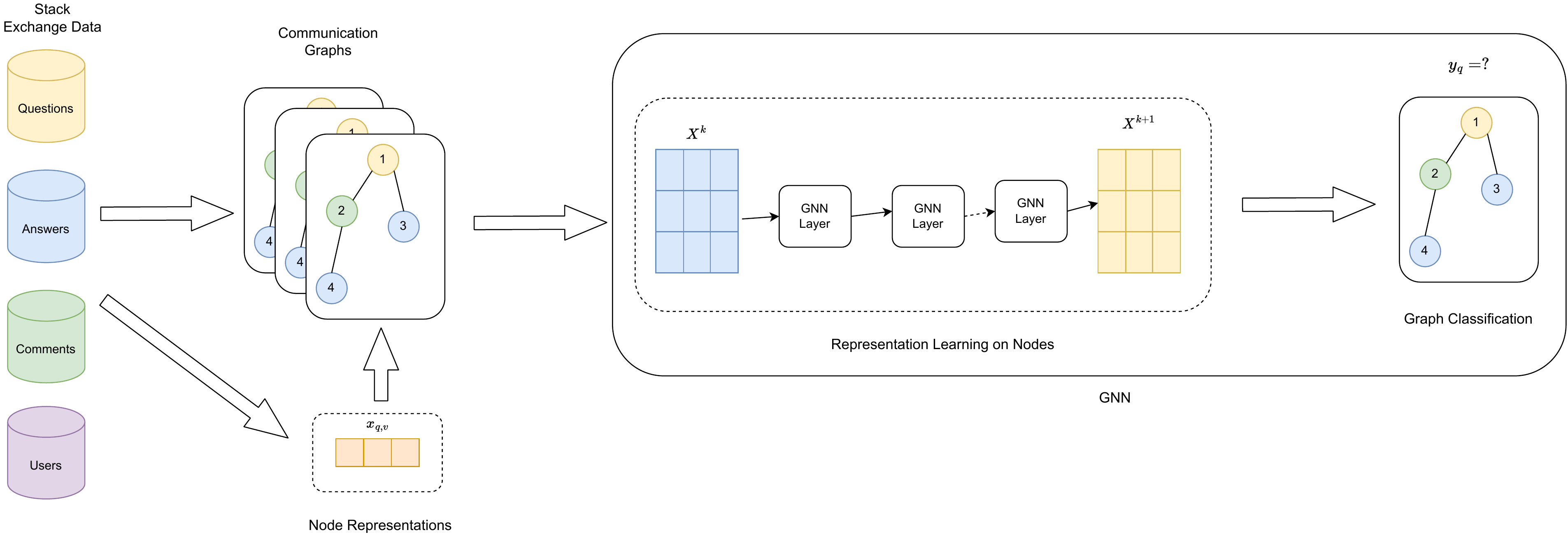}}
        \caption{Overview of the proposed approach}
        \label{fig:overview}
    \end{figure*}

    In this work, we propose a novel and reliable approach utilising the PGM and GNNs for identifying unresolved questions in Stack Exchange Q\&A communities with high accuracy. We aim to examine and study the possible causes that may lead to unresolved questions; our approach is a first step towards that goal. In our proposed approach, we first model the communication network surrounding each question using the PGM, construct a communication graph for each question, and then utilise state-of-the-art GNN-based techniques to accurately detect unresolved questions. By communication network, we mean the network of messages (mainly in the form of answers and comments) users exchange to resolve a question. Our central hypothesis is that the expressive power of GNNs makes them suitable tools for investigating the problem of unresolved questions.~\Cref{fig:overview} shows a high-level overview of the proposed approach.
    Furthermore, we conduct thorough experiments to evaluate the effectiveness of our approach in comparison to other state-of-the-art methods which do not utilise the interconnected structure of the communication network of users.

    In summary, the followings are the main contributions of our work:

    \begin{itemize}
        \item We propose a method to model the communication network formed around a question using the PGM, which can express both content (i.e., the text of messages) and the structure of the communication (i.e., communication patterns).
        \item We propose a novel approach utilising GNNs that can reliably and accurately identify and detect unresolved questions on Q\&A communities on the SE platform.
        \item We experimentally evaluate the effectiveness of our approach against the baselines on real-world historical data of three distinct Q\&A communities.
        \item We make the code and data used in our work publicly available so other researchers can reproduce and build on the work described in this article more easily (see~\Cref{data_and_code}).
    \end{itemize}

    In addition, broadly speaking, we believe that the work presented here can contribute to the following subjects:

    \begin{itemize}
        \item[-] \textit{Academic Research}. In the academic realm, understanding why certain questions remain unresolved could provide insights into knowledge gaps in specific areas. These insights, for example, could be used to guide curriculum development, direct research efforts, or inspire the creation of new courses.

        \item[-] \textit{User Experience Enhancement}. Identifying unresolved questions can be used to improve the user experience by notifying users about unresolved questions in their areas of expertise, encouraging them to contribute and enhancing their engagement on the platform.

        \item[-] \textit{Question Quality Analysis}. Over time, our model could be used to analyse the quality of questions being asked. If a significant number of questions remain unresolved, it could suggest that the questions are unclear or too broad. This information could lead to the development of guidelines or resources to help users ask better questions which in turn can improve user satisfaction.
        
    \end{itemize}

    The rest of this article is organised as follows:
    ~\Cref{related_work} describes and discusses the related work.~\Cref{preamble} introduces the essential concepts and techniques needed to understand better the methodology used in this study. \Cref{data} provides the details of the data used in the experiments.~\Cref{results} provides information about the experiments, including the information about the baselines and the evaluation metrics.~\Cref{discussion} presents the results and discusses the subsequent important findings and \Cref{future_work} discusses the limitations of our work and suggests future work. Finally,~\Cref{conclusion} concludes this article.

    \begin{figure*}[hbt!]
        \scriptsize {
            \begin{subfigure}{.47\linewidth}
                \includegraphics[width=\linewidth]{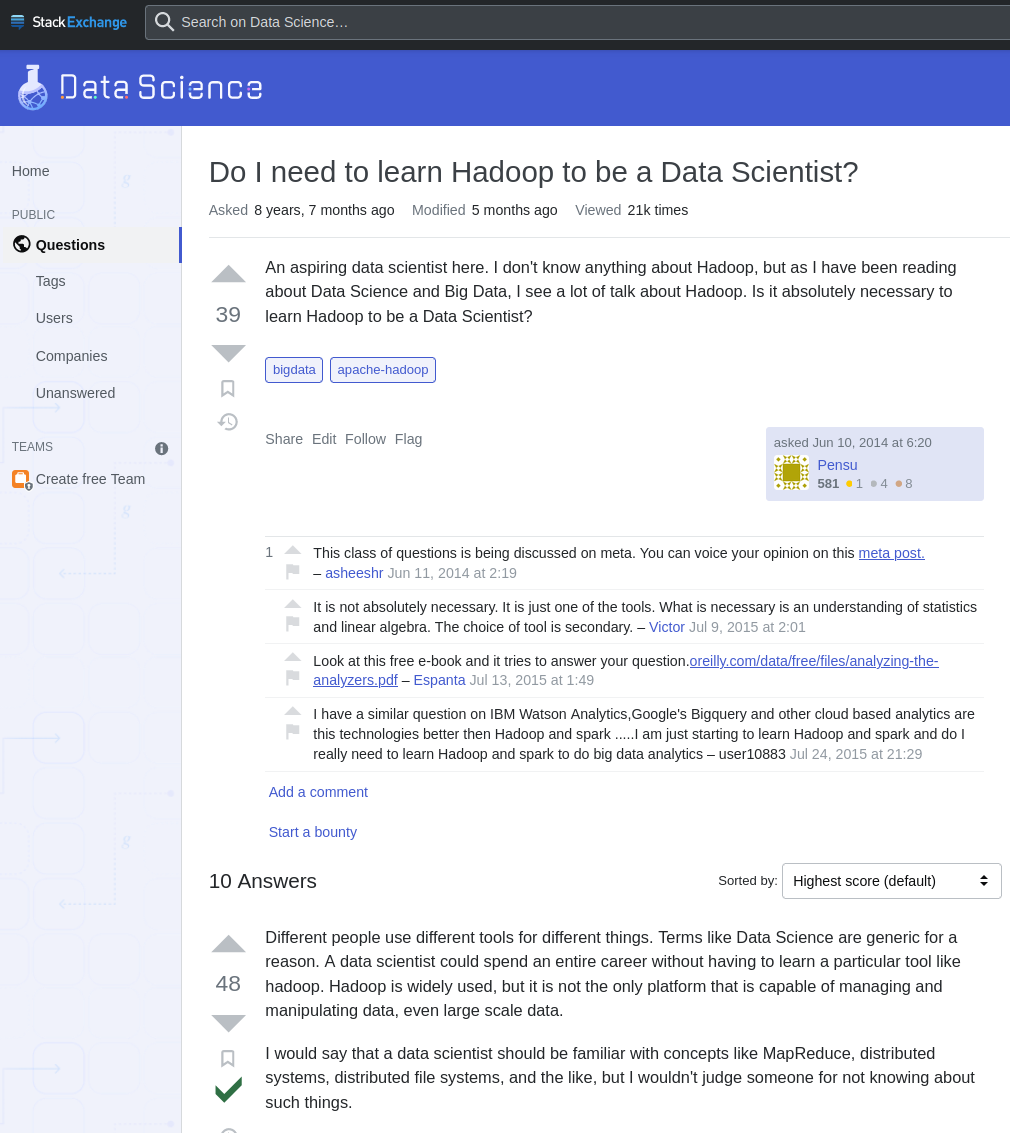}
                \caption{The actual question}
                \label{fig:ds_q253}
            \end{subfigure}\hfill 
            \begin{subfigure}{.47\linewidth}
                \includegraphics[width=\linewidth]{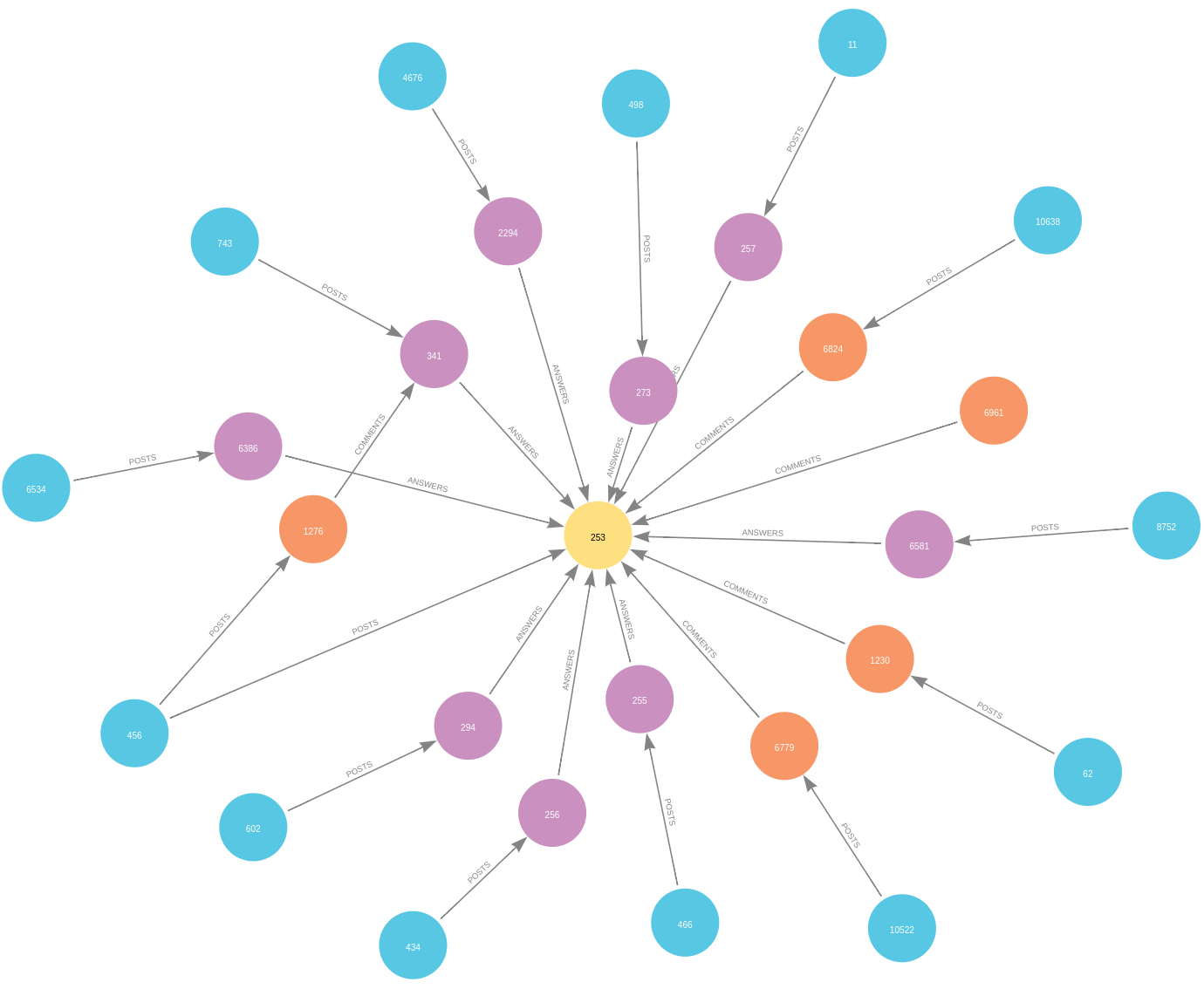}
                \caption{The communication graph of the question (with id = 253); the blue nodes represent users, the purple nodes represent answers, the orange nodes represent comments, and the yellow node in the middle represents the question. Note that the text property fields of the nodes are not shown}
                \label{fig:prop_graph}
            \end{subfigure}
            \caption{An illustrative example showing the question with id 253 from the Data Science SE community; the picture of the actual question can be seen on the left; the corresponding communication graph (which was created by modelling the corresponding communication network as a property graph) is shown on the right}
        }
    \end{figure*}

    \section{Related work}\label{related_work}

    In recent years, the interest in studying phenomena in Q\&A platforms such as Stack Exchange and Quora\footnote{\href{https://www.quora.com/}{https://www.quora.com}} has skyrocketed, as these platforms offer a user-friendly setting for sharing accessible knowledge. For example, authors in~\citep{Anderson20212} explored the foundations for understanding community processes in Q\&A platforms, using Stack Overflow as an example. Their work highlights the importance of considering the best answer to a question, the set of answers, and the processes that create them. They used the temporal structure of Stack Overflow to identify properties of questions and their answers that are likely to have lasting value and those that require further community involvement. Furthermore, authors in~\citep{Asaduzzaman2013} developed a taxonomy to understand the factors involved in questions receiving no answers on Stack Overflow. They used the taxonomy to build a classifier that predicts how long a question will remain unanswered. They found that about 7.5\% of the questions on Stack Overflow are unanswered. And it suggests that identifying the quality of a question could also help better determine whether a question will receive an answer. Similarly, authors in~\citep{Goderie2015} investigated the possibility that the time between a question is posted and when it gets a response on the Stack Overflow Q\&A community could be predicted with reasonable accuracy. They achieved an accuracy of around 30\% to 35\%.

    Furthermore, and more relevant to our work, authors in~\citep{Yazdaninia2021} developed a predictive model utilising XGBoost~\citep{Chen2016} to identify unresolved questions on Stack Overflow. Their model requires extensive feature engineering to work correctly. In contrast, our approach uses state-of-the-art text embedding methods to minimise the need for hand-crafted feature engineering, which is typically challenging.

    A very much recent development in the area of automated Q\&A is the emergence of AI-powered conversational systems such as ChatGPT~\citep{Bubeck2023}, which benefited from the recent advancement in the field of natural language processing (NLP). Although the effect of the use of systems like ChatGPT is not still well-understood due to the recency of their emergence, nonetheless, it has been a growth in the amount of attention poured into this area by scholars recently. In~\citep{Sanatizadeh2023}, authors explored the impact of large language models (LLMs), like ChatGPT and Bard, on online Q\&A platforms, specifically focusing on the Stack Exchange platform. Their study reveals two key trends: an increase in the quality and complexity of both questions and answers and a decrease in platform engagement (visits, posts, user activity), particularly in the Technology sector post-LLM introduction. Their research suggests that while LLMs enhance content depth, they may reduce overall user engagement, presenting a potential need for future research. Work presented in~\citep{Xue2023} analysed the impact of LLMs, such as ChatGPT, on Q\&A communities, using Stack Overflow as a case study. Following ChatGPT's launch, a 2.64\% average reduction in question-asking was observed, suggesting that LLMs decrease search costs, leading to fewer but potentially more engaging and higher-quality questions. However, while questions became 2.7\% longer (indicating sophistication), they were found to be less readable and cognitively challenging, possibly making them difficult for LLMs to understand and process. Further, no significant change in viewer scores suggested no improvement in question quality and decreased engagement across the platform. The study concludes that LLMs might pose a risk to the survival of Q\&A communities, potentially impacting LLMs' sustainable learning and long-term improvement, with new users being the most affected. Finally,~\citep{Sohail2023} conducted a comprehensive survey of over 100 recent publications on ChatGPT. It has demonstrated impressive achievements since its inception in November 2022 but still grapples with biases and a lack of trust. The authors propose a taxonomy for ChatGPT research, identify common methodologies, and explore its application areas and critical issues. The paper also outlines future research directions, suggests solutions to current challenges, and speculates on future advancements. It is presented as the first comprehensive review of ChatGPT and emphasises the vast potential for further research and development across diverse application areas. Overall, the emergence of ChatGPT and LLM-related technologies seems like an exciting direction for future research, which can open up new rays of insight into the working of Q\&A platforms, especially concerning user engagement and satisfaction.

    Overall, recent advancements in natural language processing have led to the development of AI-powered conversational systems, such as ChatGPT. While their impact is still being studied, research suggests that they enhance the quality and complexity of questions and answers, but may decrease user engagement. For example, a reduction in question-asking was observed following ChatGPT's launch, potentially impacting the sustainability of Q\&A communities. However, the technology's potential for future research and development across diverse application areas is vast. Overall, the emergence of ChatGPT and LLM-related technologies offer exciting possibilities for gaining new insights into Q\&A platforms and user satisfaction.

    \section{Preamble}\label{preamble}

    \subsection{Property graph model}

    In this work, we used the property graph model to represent the communication network of users. We created the corresponding communication graph of a communication network by carefully designing a general schema that could express both the content (i.e., messages exchanged by users) and the structure of communication taking place between users. Formally, the communication graph is a quintuple $G=(V, E, \mu, \lambda, \theta)$ where $V$ is the set of nodes. A node $v \in V$ can be of the following items: a question, an answer, a comment, or a user. Moreover, $E$ is the set of edges and $\mu: E \to V \times V$ is a function that assigns an ordered pair of nodes to each edge $e \in E$. And $\lambda: V \cup E \to L$ is a function that assigns each node or edge a label from the label set $L$. And $\theta: (V \cup E) \times K \to N$ is a function that assigns values to the properties assigned to each node or edge; $K$ is the set of properties that a node or an edge can have, and $N$ is the set of values that the properties can accept. \Cref{tab:commg_nodes} includes the information about the nodes, and \Cref{tab:commg_edges} presents the information about the edges in the communication graph. Furthermore, {\Cref{fig:commg} shows the schema of the communication graph. And~\Cref{tab:commgraphs_stats} presents the statistical characteristics of the set of communication graphs in each of the datasets introduced in~\Cref{data}.

    \begin{table*}[!htp]
        \centering
        \caption{Information about the nodes including their labels and the respective properties and values}\label{tab:commg_nodes}
        \scriptsize
        \begin{tabular}{llll}
            \toprule
            \textbf{Node type}        & \textbf{Label}            & \textbf{Property} & \textbf{Value}                                             \\\midrule
            \multirow{2}{*}{Question} & \multirow{2}{*}{Question} & id                & Unique question id                                         \\
            &                           & text              & Texts of the title and body of the question (concatenated) \\ \cline{1-4}
            \multirow{2}{*}{Answer}   & \multirow{2}{*}{Answer}   & id                & Unique answer id                                           \\
            &                           & text              & Text of the answer                                         \\ \cline{1-4}
            \multirow{2}{*}{Comment}  & \multirow{2}{*}{Comment}  & id                & Unique comment id                                          \\
            &                           & text              & Text of the comment                                        \\ \cline{1-4}
            \multirow{2}{*}{User}     & \multirow{2}{*}{User}     & id                & Unique user id                                             \\
            &                           & text              & Text from \textit{AboutMe} field of the user's profile     \\
            \bottomrule
        \end{tabular}
    \end{table*}

    \begin{table*}[!htp]
        \centering
        \caption{The information of the edge labels. Notice that edges do not have any properties}\label{tab:commg_edges}
        \scriptsize
        \begin{tabular}{lllll}
            \toprule
            \textbf{Edge label}       & \textbf{Source node label} & \textbf{Destination node label} & \textbf{Descritpion}           \\\midrule
            \multirow{3}{*}{POSTS}    & \multirow{3}{*}{User}      & Question                        & User posted the question       \\
            &                            & Answer                          & User posted the answer         \\
            &                            & Comment                         & User posted the comment        \\ \cline{1-4}
            ANSWERS                   & Answer                     & Question                        & Answer answered the question   \\ \cline{1-4}
            \multirow{2}{*}{COMMENTS} & \multirow{2}{*}{Comment}   & Question                        & Comment commented the question \\
            &                            & Answer                          & Comment commented the answer   \\
            \bottomrule
        \end{tabular}
    \end{table*}

    \begin{figure}[t]
        \centerline{\includegraphics[scale=0.8]{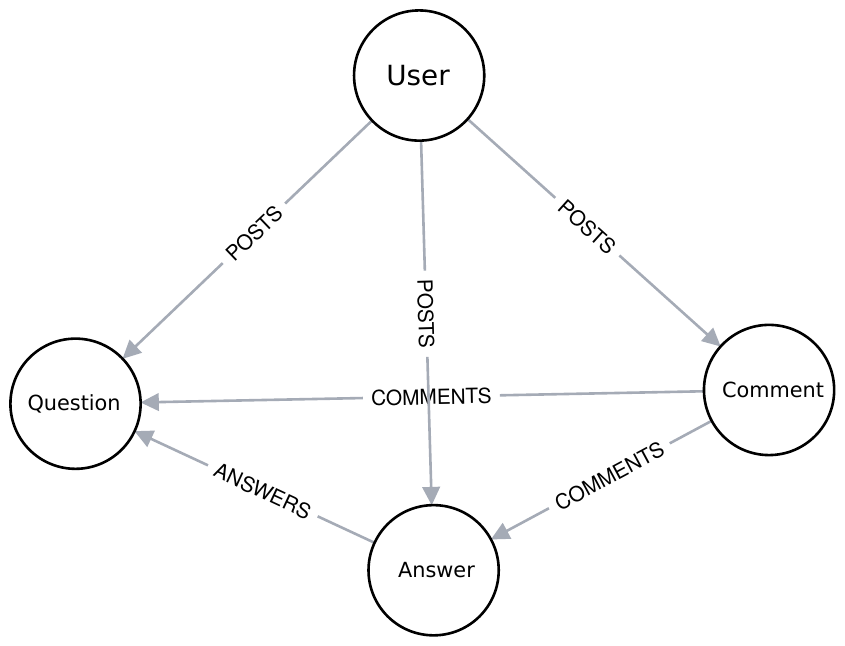}}
        \caption{The figure shows the schema of the communication graph, including the different nodes and edges distinguishable by their respective labels}
        \label{fig:commg}
    \end{figure}

    \subsection{Graph neural networks}

    Graph neural networks (GNNs) are deep neural networks that utilise the linked structure of the data they operate on. In principle, GNNs work by iteratively updating node representations. They achieve this by combining every node's representation with its neighbouring nodes' representations in each iteration. Formally, given graph $G=(V, E)$ where $V$ is the set of nodes, and $E$ is the set of edges, let $H^0$ be the set of initial node representations where $H^{0}_v$ is the initial representation of node $v$. Then, to implement a GNN over $G$, we need to devise a neural network with the following two important functions at each layer: an aggregation function and a combination function. More specifically, starting with initial node representations $H^0$, and number $k \in \{1, 2, 3, ..., K \}$ which indicates the $k$th iteration, the aggregation and combination functions will have the following signatures:

    \begin{itemize}
        \item Aggregation: $a^{k}_v = Aggregate(H^{k-1}_u)$ where $u \in N(v)$ is a neighbouring node of $v$ (i.e., $N(v)$ is the set of all nodes adjacent to $v$). And $a^{k}_v$ is an aggregate representation of the representations of all of the neighbours of $v$ in the $(k-1)$th iteration.
        \item Combination: $H^{k}_v = Combine(H^{k-1}_v, a^{k}_v)$.
    \end{itemize}

    As you can see, at iteration $k$, for each node $v \in V$, the aggregation function produces an aggregate representation (i.e., $a^{k}_v$) for the neighbours of $v$ at iteration $k-1$. Then the combination function combines this aggregate representation with the representation of $v$ at iteration $k-1$ in order to generate the updated representation for $v$. Finally, after $K$ iterations, $H^K$ would be the final node representations which can be utilised for downstream machine learning tasks such as the graph or node classification.

    Since the introduction of the GNNs by authors of~\citep{Scarselli2009}, there has been a growing interest in the GNNs from the research community, which has led to the invention of many different architectures and types of GNNs.
    One significant difference between various GNN architectures is how they define and implement their aggregation and combination functions. In this study, we utilised two types of GNN architectures, namely, graph convolutional neural networks (GCNs) ~\citep{Kipf2016}, arguably the most popular graph neural network architecture due to its simplicity and effectiveness for solving various tasks and applications~\citep{Wu2022}. And general neural networks (GGNNs)~\citep{You2020}, a more recent architecture that has shown good performance for graph classification.

    \subsection{Few-shot learning}

    Few-shot learning is a machine learning task aiming to train a classifier using a small number of labelled samples. More specifically, given $c$ classes and $k$ samples, the objective is to use $m$ samples per class, where $m << k$, to train the classifier.

    In this work, we utilised SetFit few-shot learning framework~\citep{Tunstall2022} from HuggingFace\footnote{\href{https://huggingface.co}{https://huggingface.co}}. SetFit is a state-of-the-art few-shot learning framework that uses a contrastive learning approach to fine-tune an already-trained sentence transformer model~\citep{Reimers2019}. Sentence transformers are modifications of pre-trained transformer models that use Siamese and triplet network structures to derive semantically meaningful dense sentence embeddings for input text sequences~\citep{Reimers2019}.

    Formally, for a binary classification task, the training sample set for contrastive learning is created as follows: given a small dataset of samples $D=\{(x_i, y_i)\}$, where $x_i$ is a sample, and $y_i \in \{0, 1\}$ is the corresponding label, two sets of paired samples are created. Namely, the set of positive samples $R^c$ and the set of negative samples $N^c$ where $c \in \{0, 1\}$ indicates class label. A triplet $(x_i, x_j, 1) \in R^c$ has the property that $y_i = y_j = c$ (the labels are the same and equal to $c$), and for $(x_i, x_j, 0) \in N^c$, respectively, $y_i = c$ and $y_j \neq c$ (the labels are different and only the first sample belongs to the class with label $c$). Moreover, $R^c$ and $N^c$ are constructed by randomly picking samples from each class, and $\abs{R} = \abs{N}$. Then the training set $T$ is made by concatenating the corresponding subsets of samples from the two sets as $T=\{(R^0, N^0),  (R^1, N^1)\}$. Using this method, given a small sample size, $m$, the size of the training set $T$ can be as large as ${m \times (m - 1)} \over {2}$ samples. Consequently, even a few labelled initial examples can generate enough samples for contrastive learning.

    The SetFit framework consists of two major components: a (model) fine-tuner and a classification head. The fine-tuner uses contrastive learning to fine-tune an existing sentence transformer model. The classification head uses embeddings from the fine-tuned model to classify the samples using a logistic regression learner~\citep{Yu2011}.

    In this work, we opted to use a few-shot learning-based approach, more specifically SetFit framework, as baselines since transformer-based neural networks, which power SetFit, such as BERT~\citep{Devlin2018} have shown excellent performance in applications such as text classification and embedding.

    \section{Data}\label{data}

    \subsection{Data description}

    In the work described in this article, we used the data from~\citep{AbediFirouzjaei2022}. The data include the complete historical data of three SE communities, namely, Computer Science (CS) SE, Data Science (DS) SE, and Political Science (PS) SE, from their inception up until May 2021. The main reason we chose these datasets was that they represent distinct communities of users with different interests, which allowed us to more reliably evaluate and compare the performance of our approach with other methods. \Cref{tab:desc} includes the information about the datasets used in this work. For more extensive information about the datasets, including the detailed characteristics of the communities these datasets represent, please see~\citep{AbediFirouzjaei2022}.

    \begin{table}[t]
        \begin{center}
            \caption{Information about the datasets}
            \begin{tabular}{lccc}
                \toprule
                \multirow{2}{*}{\textbf{Characteristic}} &
                \multicolumn{3}{c}{\textbf{Dataset}} \\ &
                \scriptsize{\textbf{Pol}} & \scriptsize{\textbf{DS}} & \scriptsize{\textbf{CS}} \\ \hline
                \midrule
                \scriptsize{\#Questions}              & \scriptsize{11,853}  & \scriptsize{28,768}  & \scriptsize{39,794}  \\ \hline
                \scriptsize{\#Users}                  & \scriptsize{31,242}  & \scriptsize{100,582} & \scriptsize{113,434} \\ \hline
                \scriptsize{\#Answers}                & \scriptsize{24,712}  & \scriptsize{32,107}  & \scriptsize{45,517}  \\ \hline
                \scriptsize{\#Comments}               & \scriptsize{126,838} & \scriptsize{63,677}  & \scriptsize{136,085} \\ \hline
                \scriptsize{\% of resolved questions} & \scriptsize{52\%}    & \scriptsize{34\%}    & \scriptsize{46\%}    \\ \hline
                \scriptsize{Foundation year}          & \scriptsize{2012}    & \scriptsize{2014}    & \scriptsize{2008}    \\ \hline
                \bottomrule \hline
            \end{tabular}
            \label{tab:desc}
        \end{center}
    \end{table}

    \subsection{Node representations}

    We used a state-of-the-art sentence transformer model, namely all-MiniLM-L6-v2\footnote{\href{https://huggingface.co/sentence-transformers/all-MiniLM-L6-v2}{https://huggingface.co/sentence-transformers/all-MiniLM-L6-v2}} to transform user-generated text (i.e., posts, comments, etc.) into 384-dimensional semantic embedding vectors. We used the semantic embedding vectors as features for one of the baseline methods (i.e., the logistic regression learner) and also to create node representations. In addition, specifically, to be used in our approach, we embedded the information about the node type in the communication graphs as a sparse one-shot vector. 
Furthermore, we used a binary coding scheme to represent the types of the nodes.~\Cref{tab:node_features} shows the information about the representation of the features attached to nodes in the communication graphs used in our experiments. Please, see the implementation files for more detail (see \Cref{data_and_code}).

\begin{table*}[!htp]
\begin{center}
\caption{The specification of features assigned to each type of the nodes in the communication graph}\label{tab:node_features}
\scriptsize
\begin{tabular}{lllll}\toprule
\textbf{Representation} &\textbf{} &\textbf{Format} &\textbf{Description} \\\midrule
Text embedding & & $[a_0, a_2, ..., a_{383}]$, $0 \leq a_i \leq 1$ & Semantic embedding vector of the text\\ \cline{1-4} 
\multirow{4}{*}{Node type} &Question &[0, 0, 0, 1] & Vector representation of a question node \\
&Answer &[0, 0, 1, 0] & Vector representation of an answer node\\
&Comment &[0, 1, 0, 0] & Vector representation of a comment node \\
&User &[1, 0, 0, 0] & Vector representation of a user node\\
\bottomrule
\end{tabular}
\end{center}
\end{table*}

    \begin{table*}[!htp]
        \centering
        \caption{Statistical characteristics of the communication graphs in each dataset}\label{tab:commgraphs_stats}
        \scriptsize
        \begin{tabular}{lllllllll}
            \toprule
            \textbf{Dataset}        & \textbf{Characteristic} & \textbf{\#Nodes} & \textbf{\#Edges} & \textbf{\#Answers} & \textbf{\#Comments} & \textbf{\#Users} \\\midrule
            \multirow{8}{*}{Pol SE} & Mean                     & 21.35            & 26.57            & 2.08               & 10.70               & 7.56             \\
            & Mode                     & 4                & 7                & 1                  & 2                   & 3                \\
            & Median                   & 15               & 17               & 1                  & 7                   & 6                \\
            & First Quartile           & 8                & 9                & 1                  & 3                   & 3                \\
            & Third Quartile           & 26               & 33               & 3                  & 13                  & 9                \\
            & Standard Deviation       & 21.65            & 30.24            & 2.05               & 13.48               & 6.91             \\
            & Maximum                  & 276              & 399              & 27                 & 179                 & 76               \\
            & Minimum                  & 2                & 1                & 0                  & 0                   & 1                \\
            &                          &                  &                  &                    &                     &                  \\
            \multirow{8}{*}{DS SE}  & Mean                     & 7.03             & 7.68             & 1.13               & 2.21                & 2.69             \\
            & Mode                     & 4                & 3                & 1                  & 0                   & 2                \\
            & Median                   & 6                & 5                & 1                  & 1                   & 2                \\
            & First Quartile           & 4                & 3                & 1                  & 0                   & 2                \\
            & Third Quartile           & 9                & 9                & 1                  & 3                   & 3                \\
            & Standard Deviation       & 4.95             & 7.11             & 1.09               & 3.02                & 1.72             \\
            & Maximum                  & 104              & 127              & 35                 & 52                  & 44               \\
            & Minimum                  & 2                & 1                & 0                  & 0                   & 1                \\
            &                          &                  &                  &                    &                     &                  \\
            \multirow{8}{*}{CS SE}  & Mean                     & 9.55             & 11.31            & 1.13               & 4.02                & 3.40             \\
            & Mode                     & 6                & 5                & 1                  & 0                   & 2                \\
            & Median                   & 8                & 9                & 1                  & 3                   & 3                \\
            & First Quartile           & 6                & 5                & 1                  & 1                   & 2                \\
            & Third Quartile           & 12               & 15               & 1                  & 5                   & 4                \\
            & Standard Deviation       & 6.81             & 10.18            & 1.01               & 4.57                & 2.09             \\
            & Maximum                  & 137              & 183              & 13                 & 81                  & 49               \\
            & Minimum                  & 2                & 1                & 0                  & 0                   & 1                \\
            \bottomrule
        \end{tabular}
    \end{table*}

    \section{Experiments}\label{results}

    \subsection{Methods}

    As mentioned earlier, we used two GNN architectures (or methods) in our approach: the GCN and GGNN. Furthermore, for each of the two methods, we experimented with three distinct types of node representations, including the following:

    \begin{itemize}
        \item The semantic embedding vector of the text property of the node, plus the information related to the type of the node. From now on, we refer to this type of representation as \textit{text embeddings plus node type}.
        \item Only semantic embedding vector of the text property of the node. From now on, we refer to this type of representation as (only) \textit{text embeddings}.
        \item Only, the information related to the type of the node. From now on, we refer to this type of representation as (only) \textit{node type}.
    \end{itemize}

    Furthermore, we compared the performance of the methods used in our approach with the performance of the following baseline learners:

    \begin{itemize}
        \item \textit{Logistic regression}: we used a logistic regression learner to predict whether a question is unresolved. We used the semantic embedding of the text property of the question nodes in communication graphs as features. We used the all-MiniLM-L6-v2 sentence transformer model to generate the embeddings.
        \item \textit{Few-shot learning}: as mentioned earlier, we used SetFit (with all-MiniLM-L6-v2 as the base model) to train three few-shot learning models to detect the unresolved questions.
        More specifically, we used 5, 10, and 20 shots (or samples) from each class to fine-tune each of the models as mentioned above, respectively. We used the text property of the question nodes in communication graphs as input to these models.
    \end{itemize}

    \subsection{Evaluation metrics}
    For evaluating the performance of the methods used in our approach and the baseline methods, we used the following metrics:

    \begin{itemize}
        \item \textit{Accuracy}, which is the ratio of the correct prediction to the total number of predictions made. Arguably, accuracy is the most critical measure in this work because it can express the predictive power of each method with clarity.
        \item \textit{Recall}, which is the number of true positive predictions (i.e., correctly classified positive samples) out of all positive samples in the test dataset.
        \item \textit{Precision}, which is the number of true positive predictions out of all positive predictions made by the model. A high precision value means the model avoids false positive predictions (i.e., incorrectly classified positive samples).
        \item \textit{F1-score}, which is the harmonic mean of precision and recall, balances both precision and recall. F1-score is an excellent metric to use when seeking a balance between precision and recall, as it considers both. Generally, a high f1-score value suggests the model has achieved a good balance between precision and recall.
    \end{itemize}

    \subsection{Experimental settings}

    We used stratified 5-fold cross-validation to both find the optimal values of the hyperparameters and evaluate each method's performance. We trained the two methods in our approach for 400 epochs with a batch size of 32 and a learning rate of $10^{-3}$. Moreover, we trained the learners used for few-shot learning for one epoch with a batch size of 16 and a learning rate of $10^{-5}$ to fine-tune the base model. In addition, we have shared the code we developed and used to run the experiments on GitHub.com (see~\Cref{data_and_code}). This includes the implementation of each method and the complete information about the hyperparameters used during the experiments. Furthermore, we employed the 5x2cv paired t-test~\citep{Dietterich1998}, setting a significance threshold (p-value) of 0.005 to verify the statistical significance of differences in accuracy between the proposed models and the baselines. The null hypothesis of this test assumes that the two models under comparison have identical performance. Lastly, we used a comprehensive suite of software and libraries, including Neo4J Desktop\footnote{\url{https://neo4j.com/}}, Spektral\footnote{\url{https://graphneural.network/}}, Scikit-learn\footnote{\url{https://scikit-learn.org/stable/}}, PyTorch\footnote{\url{https://pytorch.org/}}, TensorFlow\footnote{\url{https://www.tensorflow.org/}}, and SetFit\footnote{\url{https://github.com/huggingface/setfit}}, to preprocess the data and implement the code used in the experiments.

    \begin{table}[]
    \begin{center}
    \caption{The GCN model is a sequential GNN model which mainly consists of three graph convolutional layers (GConv)~\citep{Kipf2016} followed by a global average pooling layer (GlobAvgPool)~\citep{Lin2013}, excluding the input and the output layers. Furthermore, all the activation functions are of type rectified linear unit (ReLU)~\citep{Fukushima1969}. Overall, the model has relative a small number of neurons, i.e., 4.5k} 
\begin{tabular}{ll} 
 \\ \hline 
 Layer (type)                                & \#Param    \\ \hline \hline 
 layer 1 (GConv)                             & 288        \\ \hline 
 layer 2 (GConv)                             & 2080       \\ \hline 
 layer 3 (GConv)                             & 2080       \\ \hline 
 layer 4 (GlobAvgPool)                       & 0          \\ \hline 
\end{tabular}
\label{tab:model-summary}
    \end{center}
\end{table}

\begin{table}[]
\begin{center}
\caption{Similar to the GCN model, GGNN is also a sequential GNN model with all the activation functions being parametric ReLU (PReLU)~\citep{He2015}. However, the model, with over a million neurons, is much more complex in comparison. Please see~\citep{You2020} for more information, including the architectural details.} 
\begin{tabular}{ll} 
 \\ \hline 
 Layer (type)                             & \#Param    \\ \hline \hline 
 layer 1 (Concat)                    & 0          \\ \hline 
 layer 2 (GlobSum)                      & 0          \\ \hline        
 layer 3 (GenConv)                    & 132,608     \\ \hline
 layer 4 (GenConv)                    & 263,680     \\ \hline 
\end{tabular} 
\label{tab:model-summary}
\end{center}
\end{table}
     
    \section{Results and discussion}\label{discussion}

    \Cref{tab:results_p},~\Cref{tab:results_ds}, and~\Cref{tab:results_cs} present the results of the experiments on Pol SE, DS SE, and CS SE datasets, respectively. Each table reports the average values of each performance metric with its corresponding dispersion (i.e., the standard deviation value). Notice that larger values indicate better performance; the best values are shown in bold.

    Based on the results from~\Cref{tab:results_p}, the GCN method (with text embeddings plus the node type information) achieved the highest accuracy of 0.61, which is noticeably higher than that of the GCN with other node representations, and the baseline methods. Moreover, GGNN (with text embeddings plus node type information) also performed well, achieving an accuracy of 0.60 and the highest precision of 0.61 among all methods. Interestingly, GGNN (with only node type information) achieved the highest recall of 0.92, but with lower precision than the other GGNN representations. As for the baselines, the logistic regression method achieved an accuracy of 0.53, which was a relatively low accuracy, slightly higher than the majority class ratio, but it had a relatively high recall of 0.69, indicating that it was able to identify most of the positive samples. On the other hand, the few-shot learners with 5, 10, and 20 shot sizes achieved the same accuracy of 0.50 on average, which was not as good as the other methods. However, it is still noteworthy considering the low number of samples used for training. It suggests that few-shot learning may have the potential to be effective for our problem. However, larger shot sizes may be required to improve the accuracy.

    Based on the results from~\Cref{tab:results_ds}, in terms of accuracy, the GGNN method (with text embeddings plus node type information) achieved the highest score of 0.72, which is slightly better than the best-performing GCN method (with text embeddings plus node type information). On the other hand, the worst-performing methods in terms of accuracy were 10-shot and 20-shot learners, which achieved an average accuracy of only 0.49. Regarding recall, GGNN (with text embeddings plus node type information) achieved the highest score of 0.54. In contrast, logistic regression had the lowest recall score of 0.08, indicating that it could not correctly identify many positive cases. As for precision, GCN (with text embeddings) had the highest score of 0.62. In contrast, few-shot learners with different shot sizes had the lowest precision scores, indicating that they could not identify many positive cases accurately. Finally, regarding the f1-score, the best-performing method was GGNN (with text embeddings plus node type information), which scored 0.56. And the logistic regression learner was the worst-performing method, with an f1-score of only 0.14, indicating poor overall method performance.

    Based on the results from~\Cref{tab:results_cs}, the best-performing method was GGNN (with text embeddings plus node type information), achieving an accuracy of 0.70 and an f1-score of 0.71. Interestingly, GGNN (with node type information as node features) achieved a recall of 0.94, indicating that it was excellent at identifying positive examples, but has a low precision of 0.59 which suggests that although using node type information as node features could help GGNN to identify positive examples, it also might result in more false positives. The logistic regression learner achieved an accuracy of 0.56, lower than the top-performing GCN and GGNN methods. The results also show that the few-shot learners did not perform very well on the dataset, as their accuracies are around 0.50, lower than the majority class ratio on the dataset.

    Overall, the results indicate that the performance of the different methods varied depending on the evaluation metric and the feature set used. However, GGNN (with text embeddings plus node type information) generally performed well across all metrics. Moreover, including text embeddings in the node representations generally improved the performance of the GCN and GGNN methods compared to using only node type information. Additionally, our approach outperformed the baselines regarding accuracy on all three datasets (see~\Cref{data}). It is worth noting that the majority class ratios for the Pol SE, DS SE, and CS SE datasets are 0.52, 0.66, and 0.54, respectively. Despite this, our approach consistently achieved higher accuracy compared to these ratios.
    On the other hand, the baseline methods' accuracy values were below the majority class ratio in most cases. Also, as stated earlier, between the two GNN methods that we used in our approach, the GGNN's performance was better in general than GCN. This difference can be attributed to the fact that GGNN has a more complex architecture comprising more than eight layers, enabling it to capture the information of the underlying graph more effectively. In comparison, the GCN used in our implementation only has three (convolutional) layers.

    In summary, the results indicate the following key findings: i) although the performance of the different methods varied depending on the evaluation metric and the specificity of the feature set, nevertheless, our approach outperformed the baselines in terms of accuracy on all three datasets; ii) even though the majority class ratios for the three datasets were relatively high, our approach consistently achieved higher accuracy than these ratios; iii) finally, the GGNN's performance was generally better than GCN, which could be attributed to the fact that GGNN has a more complex architecture, enabling it to capture the information of the underlying graph more effectively.

    \begin{table*}[!htp]
        \centering
        \caption{Results on the Pol SE dataset (with majority class ratio 0.52)}\label{tab:results_p}
        \small
        \begin{tabular}{llllll}
            \toprule
            \textbf{Method}                      & \textbf{Node represnetation}                 & \textbf{Metric} & \textbf{Mean} & \textbf{Dispersion} \\\midrule
            \multirow{12}{*}{GCN}                & \multirow{4}{*}{Text embeddings + node type} & Accuracy        & \textbf{0.61} & 0.01 \\
            &                                              & Recall          & 0.69          & 0.04                \\
            &                                              & Precision       & 0.60          & 0.01                \\
            &                                              & F1-score        & 0.64          & 0.02                \\ \cline{2-5}
            & \multirow{4}{*}{Text embeddings}             & Accuracy        & 0.58          & 0.01                \\
            &                                              & Recall          & 0.62          & 0.02                \\
            &                                              & Precision       & 0.58          & 0.01                \\
            &                                              & F1-score        & 0.61          & 0.03                \\ \cline{2-5}
            & \multirow{4}{*}{Node type}                   & Accuracy        & 0.57          & 0.00                \\
            &                                              & Recall          & 0.65          & 0.05                \\
            &                                              & Precision       & 0.58          & 0.01                \\
            &                                              & F1-score        & 0.61          & 0.02                \\ \midrule
            \multirow{12}{*}{GGNN}               & \multirow{4}{*}{Text embeddings + node type} & Accuracy        & 0.60          & 0.01                \\
            &                                              & Recall          & 0.66          & 0.10                \\
            &                                              & Precision       & \textbf{0.61} & 0.01                \\
            &                                              & F1-score        & 0.63          & 0.04                \\ \cline{2-5}
            & \multirow{4}{*}{Text embeddings}             & Accuracy        & 0.60          & 0.01                \\
            &                                              & Recall          & 0.59          & 0.07                \\
            &                                              & Precision       & \textbf{0.61} & 0.01                \\
            &                                              & F1-score        & 0.60          & 0.03                \\ \cline{2-5}
            & \multirow{4}{*}{Node type}                   & Accuracy        & 0.59          & 0.01                \\
            &                                              & Recall          & \textbf{0.92} & 0.01                \\
            &                                              & Precision       & 0.56          & 0.01                \\
            &                                              & F1-score        & \textbf{0.70} & 0.00                \\ \midrule
            \multirow{4}{*}{Logistic regression} & \multirow{16}{*}{}                           & Accuracy        & 0.53          & 0.00                \\
            &                                              & Recall          & 0.69          & 0.02                \\
            &                                              & Precision       & 0.53          & 0.00                \\
            &                                              & F1-score        & 0.60          & 0.01                \\ \midrule
            \multirow{4}{*}{Few-shot(5)}         &                                              & Accuracy        & 0.50          & 0.01                \\
            &                                              & Recall          & 0.53          & 0.13                \\
            &                                              & Precision       & 0.51          & 0.01                \\
            &                                              & F1-score        & 0.52          & 0.05                \\ \midrule
            \multirow{4}{*}{Few-shot(10)}        &                                              & Accuracy        & 0.50          & 0.01                \\
            &                                              & Recall          & 0.54          & 0.10                \\
            &                                              & Precision       & 0.51          & 0.01                \\
            &                                              & F1-score        & 0.52          & 0.05                \\ \midrule
            \multirow{4}{*}{Few-shot(20)}        &                                              & Accuracy        & 0.50          & 0.01                \\
            &                                              & Recall          & 0.56          & 0.01                \\
            &                                              & Precision       & 0.52          & 0.01                \\
            &                                              & F1-score        & 0.54          & 0.01                \\
            \bottomrule
        \end{tabular}
    \end{table*}

    \begin{table*}[!htp]
        \centering
        \caption{Results on the DS SE dataset (with majority class ratio 0.66)}\label{tab:results_ds}
        \small
        \begin{tabular}{llllll}
            \toprule
            \textbf{Method}                      & \textbf{Node feature set}                    & \textbf{Metric} & \textbf{Mean} & \textbf{Dispersion} \\\midrule
            \multirow{12}{*}{GCN}                & \multirow{4}{*}{Text embeddings + node type} & Accuracy        & 0.71          & 0.00                \\
            &                                              & Recall          & 0.43          & 0.03                \\
            &                                              & Precision       & 0.60          & 0.01                \\
            &                                              & F1-score        & 0.50          & 0.01                \\ \cline{2-5}
            & \multirow{4}{*}{Text embeddings}             & Accuracy        & 0.70          & 0.00                \\
            &                                              & Recall          & 0.31          & 0.05                \\
            &                                              & Precision       & \textbf{0.62} & 0.02                \\
            &                                              & F1-score        & 0.41          & 0.04                \\ \cline{2-5}
            & \multirow{4}{*}{Node type}                   & Accuracy        & 0.66          & 0.01                \\
            &                                              & Recall          & 0.33          & 0.01                \\
            &                                              & Precision       & 0.49          & 0.01                \\
            &                                              & F1-score        & 0.40          & 0.01                \\ \midrule
            \multirow{12}{*}{GGNN}               & \multirow{4}{*}{Text embeddings + node type} & Accuracy        & \textbf{0.72} & 0.00 \\
            &                                              & Recall          & \textbf{0.54} & 0.09                \\
            &                                              & Precision       & 0.59          & 0.03                \\
            &                                              & F1-score        & \textbf{0.56} & 0.05                \\ \cline{2-5}
            & \multirow{4}{*}{Text embeddings}             & Accuracy        & 0.71          & 0.01                \\
            &                                              & Recall          & 0.46          & 0.02                \\
            &                                              & Precision       & 0.59          & 0.02                \\
            &                                              & F1-score        & 0.52          & 0.01                \\ \cline{2-5}
            & \multirow{4}{*}{Node type}                   & Accuracy        & 0.69          & 0.01                \\
            &                                              & Recall          & 0.34          & 0.04                \\
            &                                              & Precision       & 0.56          & 0.01                \\
            &                                              & F1-score        & 0.42          & 0.03                \\ \midrule
            \multirow{4}{*}{Logistic regression} & \multirow{16}{*}{}                           & Accuracy        & 0.66          & 0.00                \\
            &                                              & Recall          & 0.08          & 0.01                \\
            &                                              & Precision       & 0.50          & 0.02                \\
            &                                              & F1-score        & 0.14          & 0.01                \\ \midrule
            \multirow{4}{*}{Few-shot(5)}         &                                              & Accuracy        & 0.52          & 0.04                \\
            &                                              & Recall          & 0.47          & 0.11                \\
            &                                              & Precision       & 0.34          & 0.01                \\
            &                                              & F1-score        & 0.39          & 0.04                \\ \midrule
            \multirow{4}{*}{Few-shot(10)}        &                                              & Accuracy        & 0.49          & 0.03                \\
            &                                              & Recall          & 0.53          & 0.07                \\
            &                                              & Precision       & 0.34          & 0.01                \\
            &                                              & F1-score        & 0.41          & 0.02                \\ \midrule
            \multirow{4}{*}{Few-shot(20)}        &                                              & Accuracy        & 0.49          & 0.00                \\
            &                                              & Recall          & 0.49          & 0.00                \\
            &                                              & Precision       & 0.33          & 0.00                \\
            &                                              & F1-score        & 0.39          & 0.00                \\
            \bottomrule
        \end{tabular}
    \end{table*}

    \begin{table*}[!htp]
        \centering
        \caption{Results on the CS SE dataset (with majority class ratio 0.54)}\label{tab:results_cs}
        \small
        \begin{tabular}{llllll}
            \toprule
            \textbf{Method}                      & \textbf{Node feature set}                    & \textbf{Metric} & \textbf{Mean} & \textbf{Dispersion} \\\midrule
            \multirow{12}{*}{GCN}                & \multirow{4}{*}{Text embeddings + node type} & Accuracy        & 0.68          & 0.01                \\
            &                                              & Recall          & 0.64          & 0.04                \\
            &                                              & Precision       & \textbf{0.65} & 0.01                \\
            &                                              & F1-score        & 0.65          & 0.02                \\ \cline{2-5}
            & \multirow{4}{*}{Text embeddings}             & Accuracy        & 0.64          & 0.00                \\
            &                                              & Recall          & 0.54          & 0.04                \\
            &                                              & Precision       & 0.62          & 0.01                \\
            &                                              & F1-score        & 0.58          & 0.02                \\ \cline{2-5}
            & \multirow{4}{*}{Node type}                   & Accuracy        & 0.65          & 0.00                \\
            &                                              & Recall          & 0.67          & 0.01                \\
            &                                              & Precision       & 0.61          & 0.00                \\
            &                                              & F1-score        & 0.64          & 0.01                \\ \midrule
            \multirow{12}{*}{GGNN}               & \multirow{4}{*}{Text embeddings + node type} & Accuracy        & \textbf{0.70} & 0.01 \\
            &                                              & Recall          & 0.81          & 0.05                \\
            &                                              & Precision       & 0.63          & 0.02                \\
            &                                              & F1-score        & 0.71          & 0.01                \\ \cline{2-5}
            & \multirow{4}{*}{Text embeddings}             & Accuracy        & 0.68          & 0.01                \\
            &                                              & Recall          & 0.73          & 0.03                \\
            &                                              & Precision       & 0.63          & 0.01                \\
            &                                              & F1-score        & 0.68          & 0.01                \\ \cline{2-5}
            & \multirow{4}{*}{Node type}                   & Accuracy        & 0.68          & 0.00                \\
            &                                              & Recall          & \textbf{0.94} & 0.02                \\
            &                                              & Precision       & 0.59          & 0.00                \\
            &                                              & F1-score        & \textbf{0.73} & 0.01                \\ \midrule
            \multirow{4}{*}{Logistic regression} & \multirow{16}{*}{}                           & Accuracy        & 0.56          & 0.00                \\
            &                                              & Recall          & 0.32          & 0.01                \\
            &                                              & Precision       & 0.54          & 0.01                \\
            &                                              & F1-score        & 0.40          & 0.01                \\ \midrule
            \multirow{4}{*}{Few-shot(5)}         &                                              & Accuracy        & 0.50          & 0.02                \\
            &                                              & Recall          & 0.44          & 0.19                \\
            &                                              & Precision       & 0.46          & 0.01                \\
            &                                              & F1-score        & 0.43          & 0.11                \\ \midrule
            \multirow{4}{*}{Few-shot(10)}        &                                              & Accuracy        & 0.50          & 0.01                \\
            &                                              & Recall          & 0.47          & 0.06                \\
            &                                              & Precision       & 0.45          & 0.01                \\
            &                                              & F1-score        & 0.46          & 0.03                \\ \midrule
            \multirow{4}{*}{Few-shot(20)}        &                                              & Accuracy        & 0.49          & 0.00                \\
            &                                              & Recall          & 0.53          & 0.07                \\
            &                                              & Precision       & 0.45          & 0.00                \\
            &                                              & F1-score        & 0.49          & 0.03                \\
            \bottomrule
        \end{tabular}
    \end{table*}

    \section{Limitations and future work}\label{future_work}

    In this paper, we presented a GNN-based approach for predicting unresolved questions in SE Q\&A communities. However, we must acknowledge our work's primary limitation: the lack of absolute forecasting utility. More specifically, when a question is posted in an online community, the preliminary information available is its content, such as the title, body, and associated tags. In order to use a GNN-based approach, information about the structure of the communication network surrounding the question is also required. This limitation can be partially addressed using GNN architectures that operate on evolving graphs, for example, as described in~\citep{Pareja2020}. However, a content-based approach, such as the baselines used in our experiments, only requires information about the questions, making it more flexible for forecasting purposes.

    Another relevant problem is: given an unresolved question and its answer, and other information, including the structure of the communication network formed around the question, how could we utilise an ML-based approach in order to rank answers and recommend promising answers to the user who asked the question to get the question resolved.

    Despite the abovementioned limitation, our results suggest that a GNN-based approach can outperform a content-based approach for predicting unresolved questions. Further research is needed to fully explore the potential of GNNs for this task, address the limitations of our current approach, and investigate the scalability and robustness of our approach on larger datasets and in different domains.

    \section{Conclusion}\label{conclusion}

    In this work, we proposed a novel approach to identify unresolved questions on Stack Exchange question-answering communities utilising the graph structure of user communication formed around a question. Our approach models the communication network encompassing a question using the property graph model. It uses graph neural networks, which can work both on the structure of communication and the content (i.e., messages exchanged among users) to identify unresolved questions. The results of our experiments show the effectiveness of the proposed approach compared to baseline methods, which only utilise the content of questions. We believe our work is a first step towards better understanding the factors that can affect questions being unresolved in Stack Exchange communities, utilising state-of-the-art graph neural network methods.

    \section*{Declarations}\label{declarations}

    \bmhead{Funding}

    This work was carried out as part of the Trondheim Analytica project\footnote{\url{https://www.ntnu.edu/trondheimanalytica}}, supported by NTNU's Digital Transformation programme.

    \bmhead{Conflict of interest}
    I do not have any conflicts or competing interests to declare.

    \bmhead{Availability of data and materials}\label{data_and_code}
    The data used in this study are publicly available from Archive.org under Creative Commons licences. Furthermore, for reproducibility, the code and other related artefacts, such as the preprocessed version of the data used in the experiments, are also available on GitHub.com~\footnote{\url{https://github.com/habedi/GNNforUnresolvedQuestions}}.

    \bibliography{references}

\end{document}